\documentclass[conference, 10pt]{IEEEtran}
\ifCLASSINFOpdf
\else
\fi
\usepackage[cmex10]{amsmath}
\usepackage{array}
\usepackage[tight,footnotesize]{subfigure}
\usepackage{graphicx}
\usepackage{color}
\usepackage{placeins}
\usepackage{float}
\usepackage{tabularx,colortbl}
\usepackage{bm}
\usepackage{amsfonts}

\begin{document}

\newtheorem{Def}{\textbf{Definition}}
\newtheorem{Theorem}{\textbf{Theorem}}
\newtheorem{Lemma}{\textbf{Lemma}}
\newtheorem{Corollary}{\textbf{Corollary}}
%
% paper title
% can use linebreaks \\ within to get better formatting as desired
\title{Quasi-Equiangular Frames (QEFs) : \\ A New Flexible Configuration of Frames}

% author names and affiliations
% use a multiple column layout for up to three different
% affiliations
%% For several authors with only one affiliation:
%%
\author{
  \IEEEauthorblockN{Hailong Shi, Hao Zhang and Xiqin Wang}
  \IEEEauthorblockA{Department of Electronic Engineering\\
    Tsinghua University\\
    Beijing, China\\
    Email: shl06@mails.tsinghua.edu.cn, haozhang@tsinghua.edu.cn}
}
\maketitle

\begin{abstract}
%\boldmath
 Frame theory is a powerful tool in the domain of signal processing and communication. Among its numerous configurations, the ones which have drawn much attention recently are Equiangular Tight Frames(ETFs) and Grassmannian Frames. These frames both have optimality in coherence, thus bring robustness and optimal performance in applications such as digital fingerprint, erasure channels, and Compressive Sensing. However, the strict constraint on existence and construction of ETFs and Grassmannian Frames becomes the main obstacle for widespread use. In this paper, we propose a new configuration of frames: Quasi-Equiangular Frames, as a compromise but more convenient and flexible approximation of ETFs and Grassmannian Frames. We will give formal definition of Quasi-Equiangular Frames and analyze its relationship with ETFs and Grassmannian Frames. Furthermore, for popularity of ETFs and Grassmannian Frames in Compressive Sensing, we utilize the technique of random matrices to obtain a probabilistic bound for the Restricted Isometry Constant (RIC) of Quasi-Equiangular Frame. Numerical simulations are demonstrated for validation.
\end{abstract}
% IEEEtran.cls defaults to using nonbold math in the Abstract.
% This preserves the distinction between vectors and scalars. However,
% if the conference you are submitting to favors bold math in the abstract,
% then you can use LaTeX's standard command \boldmath at the very start
% of the abstract to achieve this. Many IEEE journals/conferences frown on
% math in the abstract anyway.

% keywords

% For peer review papers, you can put extra information on the cover
% page as needed:
% \ifCLASSOPTIONpeerreview
% \begin{center} \bfseries EDICS Category: 3-BBND \end{center}
% \fi
%
% For peerreview papers, this IEEEtran command inserts a page break and
% creates the second title. It will be ignored for other modes.
\IEEEpeerreviewmaketitle

\section{Introduction}
% \par The concept of frame theory\cite{christensen2002introduction} has always been a powerful tool in the domain of signal processing and communications. The main object treated in frame theory is a sequence of functions $\{f_k\}_{k \in \mathcal{I}}$ ($\mathcal{I}$ is a countable index set) from a separable Hilbert space, which forms a frame for $\mathcal{H}$, if there exist positive constants (frame bounds) A and B such that
% \begin{equation}
% A\|f\|_2^2 \leq \sum_{k \in \mathcal{I}}|\langle f, f_k \rangle|^2 \leq B \|f\|_2^2
% \end{equation}

% \par
\par The concept of frame theory\cite{christensen2002introduction} has always been a powerful tool in the domain of signal processing and communications. 
 Recently various literatures have appeared concerning Equiangular Tight Frames(ETFs)\cite{Fickus20121014} and Grassmannian Frames\cite{_notationa}\cite{StrohmerGrassmannian}. These two kinds of frames both have the minimal coherence, i. e. the maximum correlation between different frame elements reaches the minimum. Frames with minimal coherence are preferred in applications such as digital fingerprint\cite{ETFfingerprint}, erasure channels\cite{Holmes200431} and Compressive Sensing\cite{Mixon2012}\cite{Bandeira2012}. However, the strict constraint on the existence and construction of ETFs and Grassmannian Frames has become the main obstacle for widespread use\cite{StrohmerGrassmannian}\cite{Waldron2009}. Although there has been some approaches like \cite{AlterativeProj} and \cite{Abolghasemi2012999} to make numerical approximation of ETFs or Grassmannian Frames, rigorous theoretical analysis on the optimality of their performance is still lack.
 
 \par
 In this paper, we will propose a new configuration of frames, Quasi-Equiangular Frames (QEFs for abbreviation), as flexible approximation of ETFs and Grassmannian Frames. Its relationship with ETFs and Grassmannian Frames will be exploited.
 % The performance of QEFs will be analyzed rigorously. That is, Asymptotically probabilistic estimation of Restricted Isometry Constant (RIC) for QEF, which provides reconstruction guarantees and performance assessments for Compressive Sensing  \cite{Decoding}\cite{RIPImplications}\cite{AnalOMP}, will be given based on calculation using technique of random hermitian matrices(\cite{FiniteRankDeform}\cite{ErdosRenyi1}\cite{ErdosRenyi2}).
Furthermore, since the recent popularity of ETFs and Grassmannian Frames in Compressive Sensing\cite{Mixon2012}\cite{Bandeira2012}, we will use the technique of random Hermitian matrices\cite{0036-0279-66-3-R02}\cite{FiniteRankDeform}\cite{ErdosRenyi1} to derive a rigorous probabilistic bound for the Restricted Isometry Constant(RIC) of QEFs, which provides performance guarantees and assessments for sparse reconstruction\cite{Decoding}\cite{RIPImplications}\cite{AnalOMP}.
\par
The remainder of this paper is organized as follows: Section II will give the definition and the probabilistic bound of RIC for QEFs. Detailed proof for main results is given in Section III. In section IV simulation results will be proposed for verification.\par
Throughout this paper, we denote by $\mathbb{E}$, $\mathrm{Var}(\cdot)$ the expectation and variance of a random variable, respectively. The $\ell_2$ norm is denoted by $\|\cdot\|_2$, the minimum and maximum eigenvalues of a Hermitian matrix $\bm X$ are represented by $\lambda_{\min}(\bm X)$ and $\lambda_{\max}(\bm X)$, and the identity matrix with dimension $k$ is denoted by $\bm I_k$.

\section{QEFs and its Restricted Isometry Constants}

\subsection{Definition of QEFs}

First of all, we give the formal definition of QEFs:
\begin{Def}\label{eETF}
A matrix $\Phi \in \mathbb{R}^{n \times N}$ whose columns form a frame is a Quasi-Equiangular Frame, if, for some $\varepsilon > 0$,
\begin{itemize}
\item the column norms $\mu_{ii}:=\|\phi_i\|_2^2$ satisfy:
\begin{equation}\label{framenorm}
1-\varepsilon \leq  \mu_{ii} \leq 1 + \varepsilon, \quad 1 \leq i \leq N
\end{equation}
% \item The rows are equal-norm and orthogonal, which indicates tight frames;
\item The frame correlations $\mu_{ij} := \langle \bm{\phi_i,\phi_j} \rangle$
% \begin{equation}\label{muij}
% \mu_{i,j} := \langle \bm{\phi_i,\phi_j} \rangle,\quad i \neq j
% \end{equation}
satisfy
\begin{equation}\label{deviation}
\mu_{E} - \varepsilon \leq | \mu_{ij}| \leq \mu_{E} + \varepsilon ,\quad i \neq j
\end{equation}
\end{itemize}
where $\mu_{E}=\sqrt{\frac{N-n}{n(N-1)}}$ is the Welch Bound\cite{Welch}.
\end{Def}
\par
From the definition it is clear that parameter $\varepsilon$ of QEFs constrains the deviation of all the frame correlations from Welch Bound (\ref{deviation}) as well as some relaxation of the frame norms (\ref{framenorm}). While the definition of ETFs\cite{Fickus20121014}\cite{Waldron2009} and Grassmannian Frames\cite{StrohmerGrassmannian} only concerns about the coherence, i.e. the maximal frame correlations:
\begin{equation}
\mu = \max_{i \neq j}|\langle\bm{ \phi_i, \phi_j }\rangle|,
\end{equation}
 ETFs are frames with coherence achieving Welch bound\cite{Welch}
\begin{equation}\label{WelchBound}
\mu\geq \sqrt{\frac{N-n}{n(N-1)}},
\end{equation}
which indicates $|\langle \bm{\phi_i , \phi_j} \rangle|$ are equal to $\sqrt{\frac{N-n}{n(N-1)}}$ for all $i\neq j$\cite{Fickus20121014}\cite{Mixon2012}. On the other hand, Grassmannian Frames have minimal achievable coherence\cite{StrohmerGrassmannian}, which may be larger than Welch bound, for the Welch bound is not achievable for all dimensions\cite{Waldron2009}. It is obvious that ETF is a special kind of Grassmannian Frame whose coherence achieves Welch bound.
\par
\begin{figure}[htbp]
  \centerline{\subfigure(ETF){\includegraphics[width=1.2in]{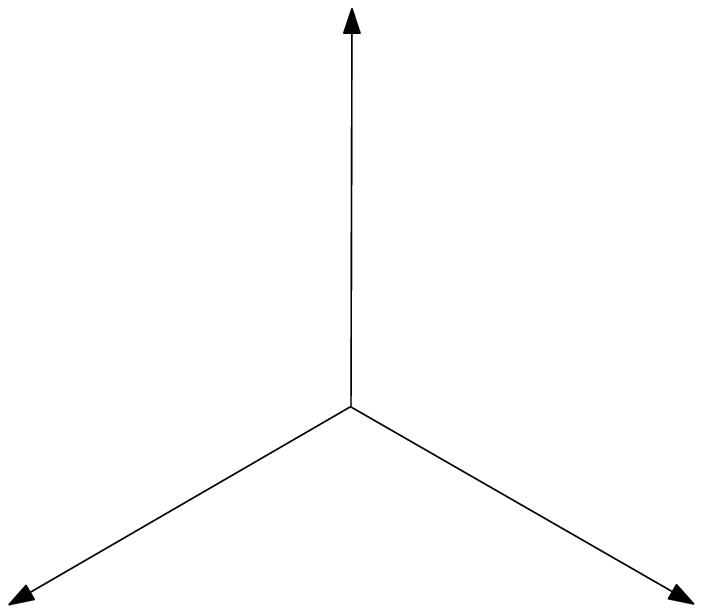}
      % where an .eps filename suffix will be assumed under latex,
      % and a .pdf suffix will be assumed for pdflatex
      \label{fig:first_case}}
    \hfil
    \subfigure(QEF){\includegraphics[width=1.2in]{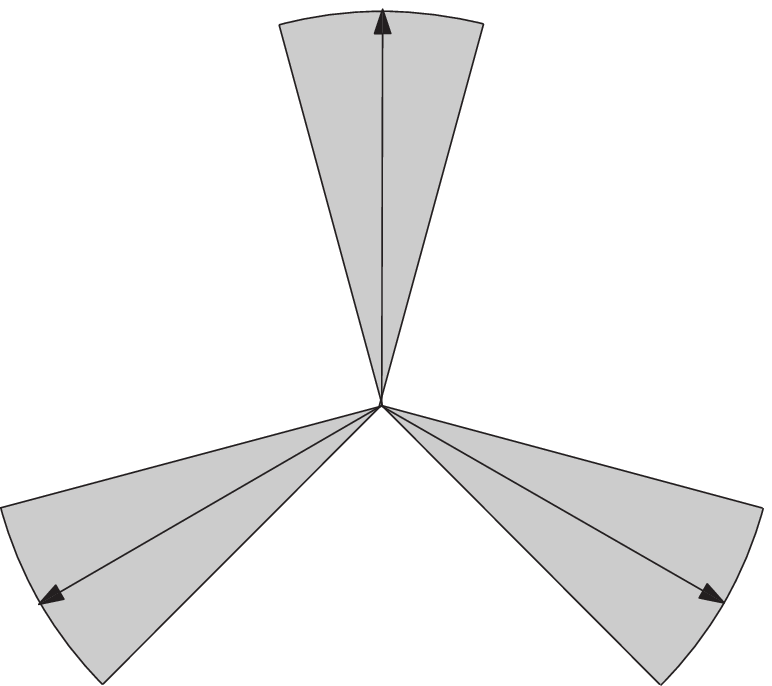}
      % where an .eps filename suffix will be assumed under latex,
      % and a .pdf suffix will be assumed for pdflatex
      \label{fig:second_case}}}
  \caption{Frame angles in 2-dimensional vector space for ETFs and QEFs}
  \label{fig:comparison}
\end{figure}

Compared with ETFs, QEFs is essentially a more flexible configuration of frames. 
Indeed, the $\varepsilon$ condition (\ref{framenorm})(\ref{deviation}) quantitatively describes the deviation of QEFs from ETFs, allowing for quantitative flexibility of QEFs. In fact, the angles between frame elements of ETFs can only take fixed values in the vector spaces with certain dimension, while QEFs have much more freedom, as is shown in Fig.\ref{fig:comparison} for a demonstration on the 2-dimensional plane.

\subsection{The Restricted Isometry Constant of QEFs}

Recently the application of ETFs in Compressive Sensing was discussed extensively(see \cite{Fickus20121014}\cite{Mixon2012}\cite{Bandeira2012} and references therein). The reason for popularity of ETFs is that sensing matrices with minimal coherence is preferred in sparse signal acquisition and reconstruction. Much attention has been drawn upon the Restricted Isometry Constants of ETFs, which is well known in the Compressive Sensing community and provides performance guarantees and assessments for sparse reconstruction\cite{Decoding}\cite{RIPImplications}\cite{AnalOMP}. The method of spectral analysis\cite{Fickus20121014}\cite{Mixon2012}\cite{Bandeira2012} has been commonly used to demonstrate the RIC of an arbitrary matrix for sparsity $k$, which says 
\begin{equation}\label{RICdef}
\delta_k = \max_{\substack{\Lambda \subset \{1,\cdots,N\} \\ |\Lambda|=k}}\|\bm \Phi_{\Lambda}^T \bm \Phi_{\Lambda} - \bm I_k\|_2,
\end{equation}
where $\bm \Phi_{\Lambda}$ denotes the sub-matrix consisting of columns of $\bm \Phi$ indexed by set $\Lambda$, and $|\Lambda|$ denotes the cardinality of the set $\Lambda$.
\par Some analysis utilized tools in graph theory\cite{Mixon2012}\cite{Bandeira2012} to demonstrate the RIC for ETFs, in which the Gram-matrix is divided as
\begin{equation}\label{graph}
\bm{\Phi^T\Phi} =\bm I + \mu_E \bm S,
\end{equation}
where $\bm S$ has zeros in the diagonal and $\pm 1$s in the off-diagonal, and can be modeled to describe the connectivity of vertices in an undirected graph (1 for non-connectivity and -1 for connectivity, \cite{Waldron2009}).
These analysis point out that an ETF with coherence $\mu_E$, has the Restricted Isometry Constant:
\begin{equation}\label{ETFRIC}
\delta_k = (k-1)\mu_E = (k-1) \sqrt{\frac{N-n}{n(N-1)}},
\end{equation}
for the sparsity level $k \leq |\Lambda|$, where $\Lambda$ denotes the set of "clique" in the graph described by $\bm S$ in (\ref{graph}). It is noted that the "clique" means the largest set of interconnected vertices in the graph described by $\bm S$, thus $\Lambda$ means the largest column set of $\bm \Phi$ such that all the off-diagonal elements of Gram-matrix $\bm{\Phi_{\Lambda}^T\Phi_{\Lambda}}$ take the value $-\mu_E$\cite{Fickus20121014}\cite{Mixon2012}.

\par
It is expected that QEFs should have approximately similar RIC to ETFs because it is some kind of approximation of ETFs as $\varepsilon$ tends to zero. As is shown in (\ref{RICdef}), estimating RIC is essentially the calculation of eigenvalues of the principal sub-matrix of the Gram-matrix. We proposed that the Gram-matrix of QEFs can be treated as random perturbations of that of ETFs, which is the technique of randomization.

\par
We adopted the theory of random matrices\cite{0036-0279-66-3-R02} to deal with the problem of estimating RIC of QEFs. More explicitly, the elements of Gram-matrix of QEFs, which are frame correlations of QEFs, are modeled as bounded random variables distributed in the intervals formulated in (\ref{framenorm})(\ref{deviation}). As the precise behavior of each correlation is unknown in practical scenario, our strategy is reasonable. We need to explore the statistical properties of eigenvalues of random Gram-matrices of QEFs. We noticed that the work of Pizzo\cite{FiniteRankDeform} and Erd\H os et al.\cite{ErdosRenyi1} both fit our quests for the eigenvalues of random Hermitian matrices with non-zero means. These results were used as the basis of our derivation.

\par
By utilizing the result of Pizzo\cite{FiniteRankDeform}, we obtained our main result for the bound of the RIC for QEFs.

\begin{Theorem}
Suppose a QEF $\bm \Phi = \{\bm \phi_i\}_{i=1}^{N} \in \mathbb{R}^{n \times N}$ is defined in Definition \ref{eETF}, if the $\mu_{ij}$'s in (\ref{framenorm}) and (\ref{deviation}) are modeled as bounded i.i.d random variables, with
\begin{equation}
\mathbb{E}(\mu_{ii}) = 1,\quad \quad \mathrm{Var}(\mu_{ii}) = \sigma^2,
\end{equation}
\begin{equation}
\mathbb{E}(\mu_{ij}) = \pm \mu_E,\quad \mathrm{Var}(\mu_{ij}) = \sigma^2,\quad i \neq j
\end{equation}
then for $\varepsilon$ sufficiently smaller than $\mu_E$, and for sparsity level $k \leq |\Lambda|$ where $\Lambda$ denotes the set of a clique, in which $\mathbb{E}(\mu_{ij}) = -\mu_E, i,j \in \Lambda,i \neq j$; the RIC of $\bm \Phi$ for sparsity $k$ satisfies
\begin{equation}
(k-1)\mu_E + \frac{\sigma^2}{\mu_E} - \frac{C \log k}{\sqrt{k}} \leq 
\delta_k \leq (k-1)\mu_E +k f + \qquad
\nonumber 
\end{equation}
\begin{equation}\label{eRIC}
%\delta_k = (k-1)\mu'_E + \frac{\sigma^2}{\mu'_E} + o(1)
% |\delta_k -[(k-1)\mu_E + \frac{\sigma^2}{\mu_E}]| \leq M [\frac{\sigma^4}{k\mu_E^3}+\frac{\sigma^3}{\sqrt{k}\mu_E^2 q}+(\log k)^\xi\sigma]
\big[ \log k+k\log\big(\frac{eN}{k}\big) + t\big]
\biggl[\frac{\varepsilon}{3} + \sqrt{\frac{\varepsilon^2}{9} + \frac{2kv}{ \log k+k\log\big(\frac{eN}{k}\big) + t}}  \biggr]
\end{equation}
with probability
\begin{equation}
\mathbb{P}  \geq 1-\exp(-t).
\end{equation}
for $t>0$ and for sufficiently large $k, n$ and $N$, where $\mu_E = \sqrt{\frac{N-n}{n(N-1)}}$, $f, v$ are parameters related with $\varepsilon$:
\begin{eqnarray}\label{Moment3}
%\mu_E = \sqrt{\frac{N-n}{n(N-1)}},\nonumber \\
f = \mathbb{E}(|\mu_{ij} - \mathbb{E}(\mu_{ij})|),v =  \mathrm{Var}(|\mu_{ij} - \mathbb{E}(\mu_{ij})|),
\end{eqnarray}
and positive constants  $C$.
\end{Theorem}
\vspace{2mm}
\par
The proof of the theorem will be given in the next section, here are some remarks.\par 
%\begin{enumerate}
%\item 
Remark 1. Just like dealing with the RIC of ETFs, we consider the sparsity level at the size of the clique. We adopted that of ETFs to the QEFs for the case that $\varepsilon$ is small sufficiently, thus we treated the clique $\Lambda$ of QEFs as similar to that of ETFs, which is the largest index set where $\mu_{ij},i,j \in \Lambda$ take the value between $-\mu_E-\varepsilon$ and $-\mu_E+\varepsilon$. There are further explorations of relationships between the size of the clique $\Lambda$ and the structure of corresponding graph, see \cite{Bandeira2012} and references therein.\par
%\item 
Remark 2.
The bound of RIC of QEFs in (\ref{eRIC}) consists of two parts, an upper bound and a lower bound. The lower bound is the probabilistic convergence result of the maximal eigenvalue in the clique, while the upper bound is derived from a concentration inequality combined with the Gershgorin's Circle theorem to bound all eigenvalues of all different sub matrices $\bm \Phi_{\Lambda}^T \bm \Phi_{\Lambda}$. (\ref{eRIC}) provides a probabilistic interval for the RIC. The upper bound is crude because of the union bound we used to deal with the joint probability in the proof. However, the result we provide is reasonable because when $\varepsilon$ tends to 0, then $f,\sigma$ and $v$ tend to 0, $\delta_k$ converges to $(k-1)\mu_E$ with probability 1, which is compatible with the RIC (\ref{ETFRIC}) of ETFs.

%% Appendix:
%% If needed a single appendix is created by
%\appendix
%% If several appendices are needed, then the command
%\appendices
%% in combination with further \section-commands can be used.
\section{The proof of Theorem 1}
%\appendix The proof of Theorem 1:\par
In order to explore the RIC of QEFs, we should consider (\ref{RICdef}) as the main problem.
% \begin{equation}\label{RICmain}
% \delta_k = \max_{\substack{\Lambda \subset \{1,\cdots,N\} \\ |\Lambda|=k}}\|\bm \Phi_{\Lambda}^T \bm \Phi_{\Lambda} - \bm I_k\|_2.
% \end{equation}
Denote by $\Lambda_0$ the index set of the clique with cardinality $|\Lambda_0|=k$, then according to Definition 1 and Theorem 1 the Gram-matrix of the QEFs indexed by the clique can be
\begin{flalign}
\bm{\Phi_{\Lambda_0}^T\Phi_{\Lambda_0}-\bm I_k} &=  \bm D_{\Lambda_0} + \bm S_{\Lambda_0} + \bm F_{\Lambda_0}& \nonumber
\end{flalign}
\begin{equation}\label{Model2}
% \lefteqn{\bm{\Phi_{\Lambda_0}^T\Phi_{\Lambda_0}-\bm I_k} =  \bm D_{\Lambda_0} + \bm S_{\Lambda_0} + \bm F_{\Lambda_0}} \nonumber \\
= \left[ \begin{array}{ccc}
				\mu_E && \\
				& \ddots & \\
				& & \mu_E
				\end{array} \right] +  
  \left[ \begin{array}{ccc}
				-\mu_E && -\mu_E\\
				& \ddots & \\
				-\mu_E& & -\mu_E
				\end{array} \right] 
				+ \{f_{i,j}\}_{i,j=1}^k,
\end{equation}
where $f_{i,j} = f_{j,i},1\leq i \leq j \leq k$ are bounded random variables with the distribution satisfying:
% \begin{eqnarray}
% f_{i,j} = \left\{ \begin{array}{ll}
					% \frac{\sqrt{\mu_E^2-\varepsilon^2}-\sqrt{\mu_E^2+\varepsilon^2}}{2} & \textrm{with probability } \frac{1}{2} \\
					% \frac{\sqrt{\mu_E^2+\varepsilon^2} - \sqrt{\mu_E^2-\varepsilon^2}}{2} & \textrm{with probability }\frac{1}{2}
					% \end{array} \right.
% f_{i,j}  \in [-\varepsilon, \varepsilon]			
% \end{eqnarray}
\begin{equation}\label{distribute}
f_{i,j}  \in [-\varepsilon, \varepsilon],\quad \mathbb{E}f_{i,j} = 0,\quad  \mathbb{E}|f_{i,j}|^2 = \sigma^2.
\end{equation}
 Then for the clique $\Lambda_0$, 
\begin{eqnarray}\label{maxeigen}
\|\bm \Phi_{\Lambda_0}^T \bm \Phi_{\Lambda_0} - \bm I_k\|_2 &=& \max_{\|\bm x\|_2=1} |\bm x^T(-\bm S_{\Lambda_0}-\bm F_{\Lambda_0}-\mu_E\bm I_k)\bm x| \nonumber \\
& \geq & \max_{\|\bm x\|_2=1} |\bm x^T(-\bm S_{\Lambda_0}-\bm F_{\Lambda_0})\bm x|-\mu_E \nonumber \\
&=& \lambda_{\max}(\mu_E \bm J - \bm F_{\Lambda_0}) - \mu_E,
\end{eqnarray}
where $\bm J$ is a matrix whose elements are all 1s, and the last equality is for the fact that when $\varepsilon << \mu_E$,
\begin{eqnarray}
k (\mu_E-\varepsilon) \leq &\lambda_{\max}(\mu_E \bm J - \bm F_{\Lambda_0})& \leq k (\mu_E+\varepsilon), \nonumber \\
-k \varepsilon \leq &\lambda_{\min}(\mu_E \bm J - \bm F_{\Lambda_0})&  \leq k \varepsilon,
\end{eqnarray}
so $\lambda_{\max}(\mu_E \bm J - \bm F_{\Lambda_0}) > |\lambda_{\min}(\mu_E \bm J - \bm F_{\Lambda_0})|$, which indicates (\ref{maxeigen}). Let
\begin{equation}
\bm A_{\Lambda_0} = -\bm S_{\Lambda_0} - \bm F_{\Lambda_0} = \mu_E \bm J - \bm F_{\Lambda_0},
\end{equation}
where $\bm F_{\Lambda_0}$ is the random Hermitian matrix described in (\ref{distribute}), then $\bm A_{\Lambda_0}$ is the random Hermitian matrix perturbed by a rank 1 deterministic matrix $\mu_E \bm J$.\par
Then we utilize Pizzo's theorem (Theorem 1.2, \cite{FiniteRankDeform}), the largest eigenvalue of $\bm A_{\Lambda_0}$ will satisfy
% \begin{equation}
% |\lambda_{\max} - \theta_1 - \frac{\sigma_0^2}{\theta_1}| \leq \frac{C \log k}{\sqrt{k}},
% \end{equation}
% where we can replace $\theta_1$ with $k\mu_E$, which is the largest eigenvalue of $\mu_E \bm J$; and $\sigma_0^2 = k \sigma$, thus we get
\begin{equation}
|\lambda_{\max}(\bm A_{\Lambda_0}) - k \mu_E - \frac{\sigma^2}{\mu_E}| \leq \frac{C \log k}{\sqrt{k}},
\end{equation}
for some positive constant $C= C(\sigma,k\mu_E)$, and for large enough $k$, with probability 1.
\par Combining (\ref{RICdef}) and (\ref{maxeigen}), we can deduce that
\begin{equation}\label{lowerbound}
\delta_k \geq \|\bm \Phi_{\Lambda_0}^T \bm \Phi_{\Lambda_0} - \bm I_k\|_2
\geq (k-1) \mu_E + \frac{\sigma^2}{\mu_E} - \frac{C \log k}{\sqrt{k}}.
\end{equation}
\par On the other side, we will apply the Gershgorin's circle theorem to obtain the upper bound of $\delta_k$.
\par For any index set $\Lambda$ with $|\Lambda|=k$, we can still decompose the Gram-matrix $\bm \Phi_{\Lambda}^T \bm \Phi_{\Lambda}$ as in (\ref{Model2}). For each eigenvalue $\lambda(\bm G_{\Lambda})$ of $\bm G_{\Lambda} = \bm \Phi_{\Lambda}^T \bm \Phi_{\Lambda} - \bm I_k$, there exists an index $i \in \Lambda$, such that
\begin{eqnarray}
|\lambda(\bm G_{\Lambda}) - f_{ii}| \leq \sum_{\substack{j \in \Lambda \\ j \neq i }} |\mu_{ij}+f_{ij}| \leq (k-1)\mu_E + \sum_{\substack{j \in \Lambda \\ j \neq i }}|f_{ij}|,
\end{eqnarray}
then we can get
\begin{equation}
|\lambda(\bm G_{\Lambda})| \leq (k-1)\mu_E + \sum_{j \in \Lambda}|f_{ij}|,
\end{equation}
% and
% \begin{equation}
% \|\bm \Phi_{\Lambda}^T \bm \Phi_{\Lambda} - \bm I_k\|_2 \leq (k-1)\mu_E + \max_{i \in \Lambda}\sum_{j \in \Lambda}|f_{ij}|,
% \end{equation}
thus the RIC will satisfy
\begin{equation}\label{equivalent}
\delta_k = \max_{|\Lambda|=k} \|\bm \Phi_{\Lambda}^T \bm \Phi_{\Lambda} - \bm I_k\|_2 \leq (k-1)\mu_E + \max_{|\Lambda|=k}\max_{\substack{i \in \Lambda}}\sum_{j \in \Lambda}|f_{ij}|,
\end{equation}
where the second term is the maximum over all different $i$'s in all index set $\Lambda$ with $|\Lambda|=k$. Since the random variables $\sum_{j \in \Lambda}|f_{ij}|$ with different $|\Lambda|$ and $i \in \Lambda$ may be highly correlated, we used the well-known Bernstein's Inequality and union bound to deduce a concentration result.\par
For concentration of sum of bounded i.i.d random variables, we have:
\begin{equation}
\mathbb{P}\{\sum_{j \in \Lambda}|f_{ij}| - kf > a\} \leq \exp\{-\frac{a^2}{2a\varepsilon/3+2kv}\},
\end{equation}
where $a>0$, $f=\mathbb{E}|f_{ij}|=\mathbb{E}(|\mu_{ij} - \mathbb{E}(\mu_{ij})|)$, and $ v=\mathrm{Var}|f_{ij}|=  \mathrm{Var}(|\mu_{ij} - \mathbb{E}(\mu_{ij})|)$ as in (\ref{Moment3}).\par
Using union bound, and combined with the stirling's formula $\log\binom{N}{k} \leq k\log\big(\frac{eN}{k} \big)$, we have
% \begin{equation}
% \mathbb{P}\{\max_{|\Lambda|=k}\max_{i \in \Lambda}\sum_{j \in \Lambda}|f_{ij}| - kf \leq a\} \geq 1 - k\binom{N}{k}\exp\{-\frac{a^2}{2a\varepsilon/3+2kv}\}
% \end{equation}
\begin{align}
&\mathbb{P}\{\max_{|\Lambda|=k}\max_{i \in \Lambda}\sum_{j \in \Lambda}|f_{ij}| - kf \leq a\} \nonumber \\
&\geq 1 - k\binom{N}{k}\exp\{-\frac{a^2}{2a\varepsilon/3+2kv}\} \nonumber \\
&\geq 1-\exp\{-\frac{a^2}{2a\varepsilon/3+2kv} + \log k+k\log\big(\frac{eN}{k}\big) \}.
\end{align}
If we let
% \begin{equation}
% 1-\exp\{-\frac{a^2}{2a\varepsilon/3+2kv} + \log k+k\log\big(\frac{eN}{k}\big) \}=  1-\exp(-t),
% \end{equation}
% then
\begin{eqnarray}
-\frac{a^2}{2a\varepsilon/3+2kv} + \log k+k\log\big(\frac{eN}{k}\big) = -t,
\end{eqnarray}
then by solving the equation above we can get
\begin{eqnarray}
\lefteqn{a = \big[ \log k+k\log\big(\frac{eN}{k}\big) + t\big]} \nonumber \\
&&\cdot \biggl[\frac{\varepsilon}{3} + \sqrt{\frac{\varepsilon^2}{9} + \frac{2kv}{ \log k+k\log\big(\frac{eN}{k}\big) + t}}  \biggr],
\end{eqnarray}
% then for integers $k$ and $N$ sufficiently large, we have $ \log [k\binom{N}{k}] + t >1$ and $ \log [k\binom{N}{k}] + t >t$, we can take
% \begin{equation}
% a^* = \log k\binom{N}{k}\Big[\frac{1}{3}\varepsilon + \sqrt{\frac{1}{9}\varepsilon^2 + 2kv}  \Big] + t \Big[\frac{1}{3}\varepsilon + \sqrt{\frac{1}{9}\varepsilon^2 + \frac{2kv}{t}}  \Big],
% \end{equation}
% so that 
% \begin{equation}
 % a^*
% &\geq& \Big( \log [k\binom{N}{k}] + t\Big)\Big[\frac{1}{3}\varepsilon + \sqrt{\frac{1}{9}\varepsilon^2 + \frac{2kv}{ \log [k\binom{N}{k}] + t}}  \Big],
% \end{equation}
% and 
% \begin{equation}
 % t \Big[\frac{1}{3}\varepsilon + \sqrt{\frac{1}{9}\varepsilon^2 + \frac{2kv}{t}}  \Big] := \tau 
% \end{equation}
then we have
\begin{flalign}
&\mathbb{P}\Big\{\max_{|\Lambda|=k}\max_{i \in \Lambda}\sum_{j \in \Lambda}|f_{ij}| \leq  kf +  \big[ \log k+k\log\big(\frac{eN}{k}\big) + t\big] &\nonumber 
%& \Big( \log k+k\log\big(\frac{eN}{k}\big) + t\Big) \Big[\frac{\varepsilon}{3} + \sqrt{\frac{\varepsilon^2}{9} + \frac{2kv}{ \log k+k\log\big(\frac{eN}{k}\big) + t}}  \Big]\Big\} \nonumber
\end{flalign}
\begin{equation}
 %\Big( \log k+k\log\big(\frac{eN}{k}\big) + t\Big) 
 \cdot \Big[\frac{\varepsilon}{3} + \sqrt{\frac{\varepsilon^2}{9} + \frac{2kv}{ \log k+k\log\big(\frac{eN}{k}\big) + t}}  \Big]\Big\} \geq 1-\exp(-t),\nonumber
\end{equation}
% \begin{flalign}
 % &\geq 1-\exp\{-t\},&
% \end{flalign}
thus we have
\begin{eqnarray}\label{upperbound}
\lefteqn{\delta_k \leq (k-1)\mu_E +k f + \big[ \log k+k\log\big(\frac{eN}{k}\big) + t\big]} \nonumber \\
&& \cdot \biggl[\frac{\varepsilon}{3} + \sqrt{\frac{\varepsilon^2}{9} + \frac{2kv}{ \log k+k\log\big(\frac{eN}{k}\big) + t}}  \biggr],
\end{eqnarray}
with probability
\begin{equation}
\mathbb{P}  \geq 1-\exp(-t).
\end{equation}
Combining (\ref{lowerbound}) with (\ref{upperbound}), we can get Theorem 1's result.

\section{Simulations}
\par
\begin{figure}[htbp]
  \centering
  \includegraphics[width=0.5\textwidth]{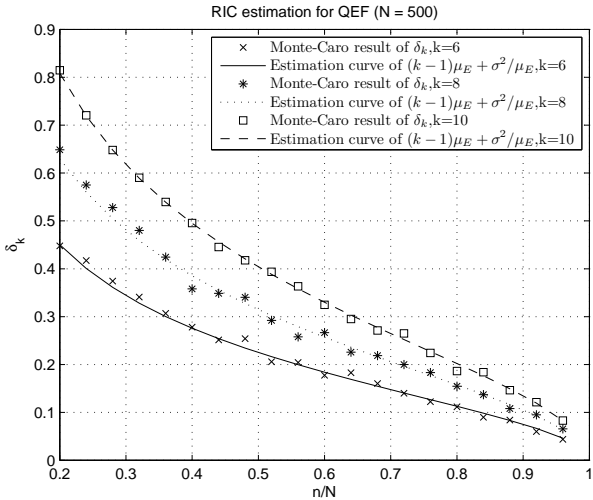}
  % where an .eps filename suffix will be assumed under latex,
  % and a .pdf suffix will be assumed for pdflatex
  \caption{Comparison between RIC lower bound (\ref{eRIC}) of QEFs and Monte-Caro result of RIC from simulations}
  \label{fig:Hermitian}
\end{figure}

In this section our bound of RIC for QEFs is verified by simulation. Unfortunately there hasn't been any explicit construction method for QEFs. In this section we only want some demonstrative validation, thus only the Gram-matrices satisfying definition 1 is randomly generated and analyzed. The dimension of the QEFs is chosen to be $N = 500, n = 100 \sim 480$ for different $\mu_E$'s, and Gram-matrix's elements $\mu_{ij},1\leq i \leq j \leq N$ is treated as uniform-distributed random variables bounded by $\varepsilon$ with mean $\mu_E$ or 1, and $\varepsilon$ is chosen to be 30\% of $\mu_E$. For the sparsity level or clique size $k$, it has been stated that finding a clique in a given graph is NP-complete, so we just build some cilques of size $k$ from 6 to 10 into our Gram-matrices. In addition, we only compare the simulated $\delta_k$ and primary part of the theoretical lower bound, which is $(k-1)\mu_E + \sigma^2/\mu_E$ in (\ref{eRIC}), because lower bound of RIC is the most concerned. Both the result of  Monte Carlo calculation and the theoretical curves are depicted in Fig. \ref{fig:Hermitian}.
It is clear that theoretical curves fits the simulation results well.
\section{Conclusion}

%The conclusion goes here.
\par
In this paper, we proposed a new configuration of frame named QEFs and discussed its relationship with other common frames including ETFs and Grassmannian Frames. QEFs provides a more flexible approximation of the ETFs and Grassmannian Frames. Theory of random Hermitian matrices and concentration inequality were utilized to derive a probabilistic bound for the Restricted Isometry Constant (RIC) of QEFs, in correspondence with parameters related to $\varepsilon$. Monte Carlo simulation was conducted to verify the correctness of our results.

%% Use \section* for acknowledgement
\section*{Acknowledgment}

This work was supported in part by the National Basic Research Program of China (973 Program, No. 2010CB731901) and in part by the National Natural Science Foundation of China (No. 40901157 and No. 61201356).

\bibliographystyle{IEEEtran}
\bibliography{CompressiveSensingMatrix.bib}
% trigger a \newpage just before the given reference
% number - used to balance the columns on the last page
% adjust value as needed - may need to be readjusted if
% the document is modified later
%\IEEEtriggeratref{8}
% The "triggered" command can be changed if desired:
%\IEEEtriggercmd{\enlargethispage{-5in}}

% references section

% can use a bibliography generated by BibTeX as a .bbl file
% BibTeX documentation can be easily obtained at:
% http://www.ctan.org/tex-archive/biblio/bibtex/contrib/doc/
% The IEEEtran BibTeX style support page is at:
% http://www.michaelshell.org/tex/ieeetran/bibtex/
%\bibliographystyle{IEEEtran}
% argument is your BibTeX string definitions and bibliography database(s)
%\bibliography{IEEEabrv,../bib/paper}
%
% <OR> manually copy in the resultant .bbl file
% set second argument of \begin to the number of references
% (used to reserve space for the reference number labels box)
% \begin{thebibliography}{1}

% \bibitem{IEEEhowto:eason}
% G. Eason, B. Noble, and I. N. Sneddon, ``On certain integrals of Lipschitz-Hankel type involving products of Bessel functions'', \emph{Phil. Trans. Roy. Soc. London}, vol. A247, pp. 529-551, April 1955.

% \bibitem{IEEEhowto:maxwell}
% J. Clerk Maxwell, \emph{A Treatise on Electricity and Magnetism, 3rd ed., vol. 2}. Oxford: Clarendon, 1892, pp. 68-73.

% \bibitem{IEEEhowto:doe}
% J. Doe, ``Title of paper if known'', unpublished. 
% \end{thebibliography}

% that's all folks
\end{document}